# Infrared absorption in a quantum wire in the presence of spin-orbit coupling: a technique to measure different types of spin-orbit interaction strengths

S. Bandyopadhyay[1] and S. Sarkar[2]

Department of Electrical and Computer Engineering

Virginia Commonwealth University

Richmond, Virginia 23284, USA

We show that the dominant absorption peak due to inter-subband transition in a gated quantum wire, with two occupied subbands, will split into a main peak and two satellite peaks if *both* Rashba and Dresselhaus spin-orbit interactions are present. One satellite peak will be red-shifted, and the other blue-shifted. From the relative intensity of either satellite peak, and the magnitude of the red- or blue-shift, we can determine *both* Rashba and Dresselhaus interaction strengths *separately*, if we also carry out a Hall measurement to determine the carrier concentration and a quantized conductance step measurement to determine the energy separation between subbands. This method may be a convenient alternative to usual magneto-transport measurements used to measure spin orbit interaction strengths. It is also more powerful because it allows us to measure the strengths of the two types of interactions separately.

PACS: 78.67.Lt, 73.21.Hb, 78.30.Fs, 71.70.Ej

---

[1] Corresponding author. E-mail: sbandy@vcu.edu
[2] On leave from Department of Electronics and Telecommunication Engineering, Jadavpur University, Kolkata, India

Measuring the spin-orbit interaction strength in semiconductor materials is an important objective since this interaction forms the basis of many spintronic devices [1] and even quantum computers [2]. Normally, the interaction strength is measured using magneto-transport techniques, such as beating patterns in Shubnikov-deHaas oscillations [3]. Such experiments are difficult, and sometimes inconclusive. First, the material must have a relatively high mobility for the oscillations to be visible at reasonable magnetic field strengths. Second, considerable theoretical analysis is required to extract the strength of the interaction from the beating patterns [3]. Finally, such beating patterns need not necessarily originate from spin-orbit interaction at all. In fact, they could be due to magneto-intersubband scattering [4]. In view of all this, there is a demand for alternate unambiguous techniques to measure the spin-orbit interaction strength in materials.

In this letter, we present an alternate technique that can be used to measure the spin-orbit interaction strength in a quantum wire using optical (infrared) absorption, preceded by a simple Hall (or Shubnikov-deHaas) measurement to determine the carrier concentration. A quantized conductance step experiment [5, 6] will also be necessary to determine the inter-subband spacing in energy, as well as ensure that only the two lowest subbands are occupied by carriers. Our technique is different from other optical techniques such as the circular photo-galvanic effect, which can measure spin orbit interaction strength, but requires monitoring photocurrents [7]. Therefore, that method is applicable only to materials that have sufficiently long radiative recombination lifetimes to produce enough photocurrent. Our method does not suffer from such shortcomings. Moreover, it allows separate determination of the Rashba and Dresselhaus interaction strengths, which none of the other methods allow. We will illustrate our technique for a quantum wire structure since that is the preferred structure for popular spintronic devices based on manipulating the spin-orbit interaction strength in materials [1].

Consider a quantum wire structure defined by split gates on a two dimensional electron gas, as shown in Fig. 1. The split gates induce a parabolic confinement on the electron gas, forming a quantum wire. We will assume that the quantum wire is in the [100] crystallographic direction. To the lowest order, the energy dispersion relations of spin-split subbands in the quantum wire are given by [8]:

$$E_n^\pm = E_\Delta + (n+1/2)\hbar\omega + (\hbar^2/2m^*)[k^2 \pm 2(m^*/\hbar^2)(\eta^2 + \alpha_n^2)^{1/2}k] = E_n + (\hbar^2/2m^*)[k \pm \kappa_n]^2$$
$$\kappa_n = (m^*/\hbar^2)(\eta^2 + \alpha_n^2)^{1/2} \qquad (1)$$
$$E_n = E_\Delta + (n+1/2)\hbar\omega - (\hbar^2/2m^*)\kappa_n^2$$

where the '+' sign (in the ±) refers to one spin and the '-' sign refers to the other (orthogonal) spin. Here, $n$ is the transverse subband index, $E_\Delta$ is the confinement energy due to the (usually triangular) confining potential in the direction perpendicular to the hetero-interfaces (y-direction in Fig. 1), $\omega$ is the curvature of the confining potential due to the split-gates (i.e. confining potential in the z-direction), $k$ is the wavevector along the axis of the wire (along the x-direction), $m^*$ is the effective mass, $\eta$ is the strength of the Rashba interaction and $\alpha_n$ is the strength of the Dresselhaus interaction in the $n$-th transverse subband. The Rashba interaction is caused by structural inversion asymmetry due to the triangular potential well confining carriers in the plane of the electron gas, while the Dresselhaus interaction is caused by bulk inversion asymmetry in non-centrosymmetric materials. The Dresselhaus interaction, unlike the Rashba interaction, depends on the electron wavefunctions and is therefore different in different subbands [8]. The dispersion relations are shown schematically in Fig. 2.

The spin eigenstates in two spin-split subbands with the same index *n*, but with orthogonal spins are [8]:

$$\psi_n^+ = \phi_n(z) \begin{bmatrix} \cos\theta_n \\ \sin\theta_n \end{bmatrix} e^{ik_n x}$$

$$\psi_n^- = \phi_n(z) \begin{bmatrix} \sin\theta_n \\ -\cos\theta_n \end{bmatrix} e^{ik_n x} \qquad (2)$$

$$\theta_n = (1/2)\arctan(\alpha_n/\eta)$$

where $\phi_n(z)$ is the z-component of the wavefunction (simple harmonic oscillator wavefunction) in the *n*-th subband. Note that $\theta_n$ is subband dependent since $\alpha_n$ is subband dependent. As a result, two spin states within the same subband are orthogonal, but within two different subbands are never completely orthogonal since $\theta_p \neq \theta_q$. In other words, the spin quantization axes in different transverse subbands are only *nearly* parallel or *nearly* anti-parallel. This is entirely a consequence of the fact that the Dresselhaus interaction is subband-dependent.

The intensity of the absorption peak corresponding to excitation from the *p*-th subband to the *q*-th subband is proportional to the square of the overlap between the wavefunctions in these subbands. Since there are two non-degenerate spin states in each subband, there can be at most 2 x 2 = 4 distinct absorption peaks. Two of them will involve transitions between states with nearly parallel spins and the other two between nearly anti-parallel spins. These intensities are given by:

$$I_{p-}^{q-} = I_{p+}^{q+} = \Xi |<\phi_q(z)|\phi_p(z)>|^2 \cos^2(\theta_p - \theta_q)$$
$$I_{p-}^{q+} = I_{p+}^{q-} = \Xi |<\phi_q(z)|\phi_p(z)>|^2 \sin^2(\theta_p - \theta_q)$$
(3)

where $\Xi$ is some constant. The first line of Equation (3) corresponds to transitions between nearly parallel spin states and the second line between nearly anti-parallel spin states. The "nearly parallel" transition gives rise to the main peak and the "nearly anti-parallel" transitions cause the satellite peaks because their intensities are much weaker than the main peak intensity when $\theta_p \cong \theta_q$. From Equation (3), we see that the ratio of the intensity of a satellite peak to that of the main peak is $tan^2(\theta_p - \theta_q)$. In most cases of interest, such as in InAs wires, the Rashba interaction strength will be much larger than the Dresselhaus interaction strength [3], so that $\alpha_n << \eta$. Therefore, $tan^2(\theta_p - \theta_q) \cong [(\alpha_p - \alpha_q)/2\eta]^2 = [a_{42}m^*\hbar\omega/(2\eta\hbar^2)]^2$, where $a_{42}$ is the material constant determining the strength of the Dresselhaus interaction. Consequently, the ratio of the two intensities is

$$\frac{I_{satellite}}{I_{main}} = \frac{I_{p-}^{q+}}{I_{p-}^{q-}} = \left[\frac{a_{42}m^*(q-p)\hbar\omega}{2\eta\hbar^2}\right]^2 \quad \text{if } \alpha_n << \eta. \quad (4)$$

It is interesting to note that if the Rashba interaction strength $\eta = 0$, then $\theta_p = \theta_q = \pi/4$, and the intensity of the satellite peaks is exactly zero. Similarly, if the Dresselhaus interaction is either totally absent or equal in different subbands ($\alpha_p = \alpha_q$), then $\theta_p = \theta_q$, and again the satellite peak intensity is zero. These conditions merely reflect the fact that transitions between strictly anti-parallel spin states are forbidden since these states are mutually orthogonal. Therefore, *both* Rashba and Dresselhaus interactions are required, and the Dresselhaus interaction must be *different* in different subbands, in order for the satellite peaks to appear. The variation of spin-orbit interaction among different subbands was also recently shown to give rise to a double peak structure in the optical spectra of quantum cascade lasers [9].

Let us now consider a quantum wire in which the two lowest subbands are occupied at low temperatures. Occupancy of only the two lowest subbands in a split gate structure can be ensured by using a back gate to vary the carrier concentration in the wire [10] and monitoring the longitudinal resistance. If the quantum wire is sufficiently clean, the conductance will exhibit the well-known quantized steps as the backgate potential is varied [5]. When only two subbands are occupied, the conductance of the wire will be ~ $4e^2/h$. The Fermi level $E_F$ will then be located between the second and third subband bottom as shown in Fig. 2. Since all states below the Fermi level are filled with electrons, the lowest state to which an electron can be photoexcited is at the Fermi level. Let the Fermi wavevectors in the spin-split second subband be called $k_{F2}^+$ and $k_{F2}^-$. Because of spin splitting, $k_{F2}^+ \neq k_{F2}^-$ (see Fig. 2) and the difference between them is $2\kappa_2$.

In a quantum wire, absorptions involving third and higher subbands are increasingly weak since carrier confinement in the higher subbands gets progressively weaker. Therefore, we need to concern ourselves only with the lowest two subbands in calculating the dominant absorption characteristics. The photon frequencies corresponding to transitions between nearly parallel states (main absorption peaks) are given by

$$h\nu_{1-}^{2-} = \hbar\omega - (\hbar^2/m^*)k_{F2}^-[\kappa_2 - \kappa_1]$$
$$h\nu_{1+}^{2+} = \hbar\omega + (\hbar^2/m^*)k_{F2}^+[\kappa_2 - \kappa_1]$$

(5)

whereas those due to the "nearly anti-parallel" transitions (satellite peaks) are given by

$$h\nu_{1-}^{2+} = \hbar\omega + (\hbar^2/m^*)k_{F2}^+(\kappa_1 + \kappa_2)$$
$$h\nu_{1+}^{2-} = \hbar\omega - (\hbar^2/m^*)k_{F2}^-(\kappa_1 + \kappa_2)$$
(6)

The frequency shift between the two main peaks is therefore $h\nu_{1+}^{2+}$ - $h\nu_{1-}^{2-} = (\hbar^2/m^*)[k_{F2}^+ + k_{F2}^-][\kappa_2 - \kappa_1] \approx [k_{F2}^+ + k_{F2}^-](\alpha_2^2 - \alpha_1^2)/\eta \approx [k_{F2}^+ + k_{F2}^-]a_{42}m^*\omega/(\hbar\eta)$. We will show later that for reasonable carrier concentrations, this frequency difference is a few µeV in energy. Therefore, thermal broadening at any practical temperature will make the two main peaks overlap and appear as a single peak.

From Equations (5) and (6), we see that one satellite peak is *blue-shifted* from the main peak by an amount of energy $\Delta E_{blue} = (\hbar^2/m^*)k_{F2}^+(\kappa_1 + \kappa_2) = k_{F2}^+\left[(\eta^2 + \alpha_1^2)^{1/2} + (\eta^2 + \alpha_2^2)^{1/2}\right]$, while the other is red-shifted by an amount of energy $\Delta E_{red} = (\hbar^2/m^*)k_{F2}^-(\kappa_1 + \kappa_2) = k_{F2}^-\left[(\eta^2 + \alpha_1^2)^{1/2} + (\eta^2 + \alpha_2^2)^{1/2}\right]$. These shifts depend on $k_{F2}^+$ and $k_{F2}^-$.

Since spin splitting is small, $k_{F2}^+ \cong k_{F2}^- = k_{F2}$. The question now is how do we know $k_{F2}$? This requires performing a Hall measurement to determine the total carrier concentration $n_l$ in the quantum wire. Let $k_{F1}$ be the Fermi wavevector in the first subband, i.e. $k_{F1}$ is the wavevector where the first subband parabola intersects the Fermi level $E_F$. The two horizontally displaced parabolas in the first subband will intersect the Fermi level at slightly different wavevectors, but we will ignore that difference since the spin splitting is small. The carrier concentration in the first subband is then $n_{l1} = 2\pi(k_{F1} - \kappa_1)$, while in the second subband it is $n_{l2} = 2\pi(k_{F2} - \kappa_2)$.

From the dispersion relations (Fig. 2), we see that: $\frac{\hbar^2}{2m^*}(k_{F1} - \kappa_1)^2 = \hbar\omega + \frac{\hbar^2}{2m^*}(k_{F2} - \kappa_2)^2$.

Therefore the total carrier concentration is $n_l = n_{l1} + n_{l2} = 2\pi[\frac{2m^*}{\hbar^2}(\hbar\omega + \frac{\hbar^2}{2m^*}(k_{F2} - \kappa_2)^2)]^{1/2} + 2\pi(k_{F2} - \kappa_2)$.

We will consider a material like InAs, where the Rashba interaction is overwhelmingly dominant over the Dresselhaus interaction [3]. Therefore, $\kappa_2 \cong (m/\hbar^2)\eta$. Consequently, the total carrier concentration can be written as

$$n_l = n_{l1} + n_{l2} \approx 2\pi[\frac{2m^*}{\hbar^2}(\hbar\omega + \frac{\hbar^2}{2m^*}(k_{F2} - m^*\eta/\hbar^2)^2)]^{1/2} + 2\pi(k_{F2} - m^*\eta/\hbar^2) \quad (7)$$

Similarly, if the Rashba interaction is dominant, then the red- and blue-shifts can be re-written as

$$\Delta E_{blue} = \Delta E_{red} \approx 2k_{F2}\eta \quad (8)$$

Equations (4), (7) and (8) are three equations with four unknowns: $a_{42}$ (the strength of the Dresselhaus interaction), $\eta$ (the strength of the Rashba interaction), $k_{F2}$ and $\hbar\omega$. The last quantity can be found from quantized conductance step measurements [5, 6]. Therefore, solutions of the three simultaneous equations with three unknowns allow us to determine the strengths of the Rashba and Dresselhaus interaction strengths – $\eta$ and $a_{42}$ - *separately*.

We now proceed to estimate the magnitude of the red- or blue-shifts $\Delta E$ in order to examine if it is observable under normal conditions. Typically, $\eta \sim 10^{-12}$ eV-m in materials like InAs [3] and $\hbar\omega = 10$ meV [6]. Note that in order for $(k_{F2} - \kappa_2)$ to be positive, $n_l > 2\pi(2m^*\omega/\hbar)^{1/2} = 5.6$ x $10^8$/m in InAs. That means this is the minimum carrier concentration required for two subbands to be occupied. Let us assume $n_l \sim 10^9$/m (this can be adjusted by the back gate). This yields $k_F \sim 1.57$ x $10^9$ m$^{-1}$. Therefore, Equation (8) yields that $\Delta E_{blue} = \Delta E_{red} \sim 1.57$ meV. Infrared spectral line widths of 1.2 meV in GaAs quantum wires at a temperature of 8 K have been reported in ref. 11. Thus, the inhomogeneous or homogeneous broadening of the main peak will not interfere with the satellite peak, and the blue or red-shift will be measurable.

Finally, we need to estimate the relative height of the satellite peak with respect to the main peak in order to assess whether the satellite peak height will be above the noise floor. Equation (4) gives this relative height. The quantity $a_{42}$ was calculated to be 3 x $10^{-29}$ eV-m$^3$ in GaAs [12]. Using this value, we determine that the ratio of the heights of the satellite peak and the main peak is ~ 0.36%, which should be well above the noise floor in low temperature measurements. Thus, we believe that the effect predicted in this letter is easily observable at low temperatures.

In conclusion, we have shown that the simultaneous presence of Rashba and Dresselhaus interactions in a quantum wire splits the dominant infrared absorption peak into a main peak and two satellite peaks. The main peak is also slightly split, but from Equation (5), this splitting is ~

$(k_{F2}/\eta)[a_{42}m^*\hbar\omega/(\hbar^2)]^2 \sim 20$ μeV, which is not resolvable. The blue- or red-shifts of the satellite peaks, and their relative intensities, allow us to determine the Rashba and Dresselhaus spin orbit interaction strengths separately, if we carry out a Hall measurement and a quantized conductance step measurement to determine the carrier concentration and the subband spacing in energy.

**Figure captions**

1. The quantum wire structure defined by split gates on a two-dimensional electron gas (2-DEG). A backgate is used to vary the carrier concentration in the channel. The quantum wire axis is assumed to be along the [100] crystallographic direction.

2. The energy dispersion of spin split subbands in a quantum wire. The various quantities referred to in the text are shown. We show transitions corresponding to the two main peaks (transitions involving nearly parallel spin states) and a transition corresponding to the blue-shifted satellite peak (involving nearly anti-parallel spin states). They are labeled "Main" and "Sat". This diagram is not to scale.

3. Schematic representation of the infrared absorption spectra in a quantum wire in the presence of strong Rashba and Dresselhaus spin-orbit interactions. The main peak and the satellite peak heights are not to scale. We show the line shapes corresponding to the $E^{-0.5}$ energy dependence of the density of states in a quantum wire.

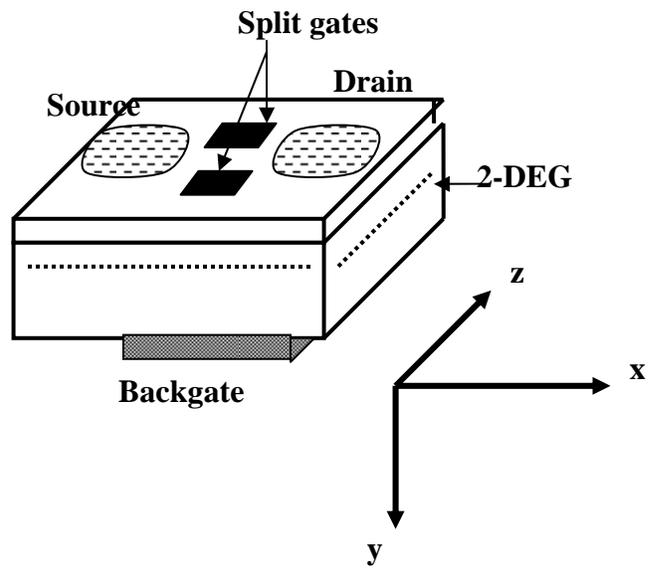

Fig. 1

Bandyopadhyay and Sarkar

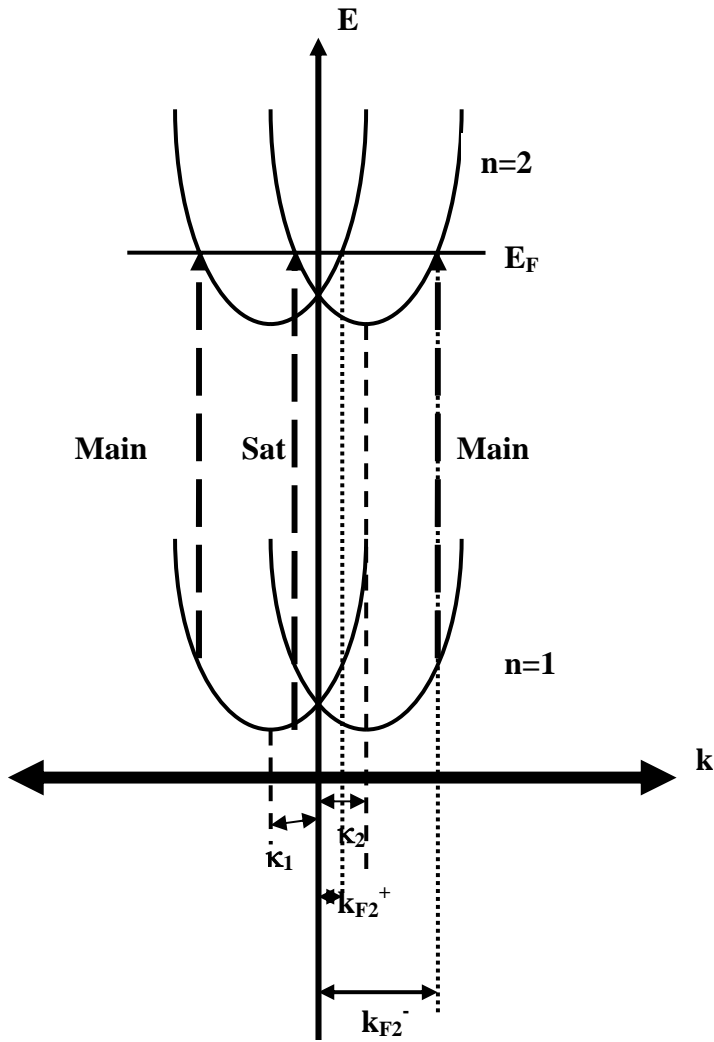

Fig. 2

Bandyopadhyay and Sarkar

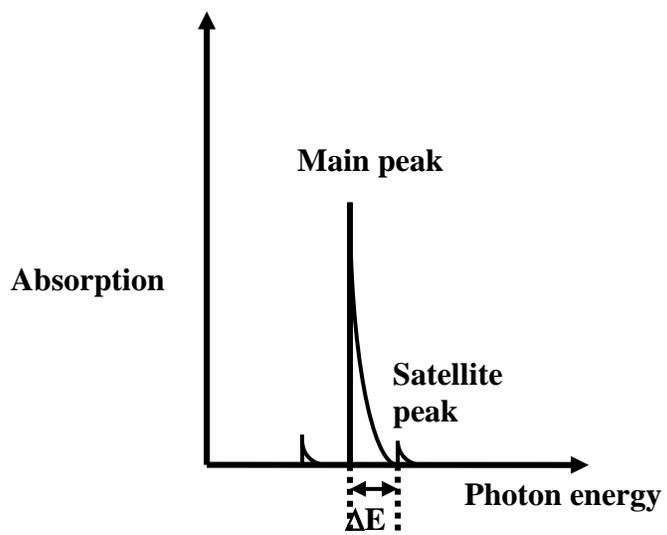

Fig. 3